\documentclass[twocolumn]{aastex62}

\usepackage{apjfonts}
\usepackage[T1]{fontenc}
\usepackage{color}
\usepackage{amsmath,amstext}
\usepackage{hyperref}
\usepackage{natbib}
\usepackage{graphicx}
\usepackage{gensymb}

\begin{document}

\title{\sc Prompt Emission of Gamma-Ray Bursts in the High-density Environment of Active Galactic Nuclei Accretion Disks}

\author{Davide Lazzati}
\affil{ Department of Physics, Oregon State University, 301
  Weniger Hall, Corvallis, OR 97331, USA}

\author{Gustavo Soares} 
\affil{ Department of Physics, Oregon State University, 301
  Weniger Hall, Corvallis, OR 97331, USA}

\author{Rosalba Perna} 
\affil{Department of Physics and Astronomy, Stony Brook
  University, Stony Brook, NY 11794-3800, USA}
\affil{Center for Computational Astrophysics, Flatiron Institute, New York, NY 10010, USA}

\begin{abstract}
Long and short gamma-ray bursts are traditionally associated with
galactic environments, where circumburst densities are  small or
moderate (few to hundreds of protons per cubic cm). However, both are
also expected to occur in the disks of Active Galactic Nuclei, where the
ambient medium density can be much larger. In this work we study, via
semi-analytical methods, the propagation of the GRB outflow, its
interaction with the external material, and the ensuing prompt
radiation. In particular, we focus on the case in which the external
shock develops early in the evolution, at a radius that is smaller than
the internal shock one. We find that bursts in such high density
environments are likely characterized by a single, long emission episode
that is due to the superposition of individual pulses, with a
characteristic hard to soft evolution irrespective of the light curve
luminosity. While multi-pulse light curves are not impossible, they
would require the central engine to go dormant for a long time before
re-igniting. In addition, short GRB engines would produce bursts with
prompt duration that would exceed the canonical 2~s separation threshold
and would likely be incorrectly classified as long events, even
though they would not be accompanied by a simultaneous supernova. 
Finally, these events have a large dynamical efficiency which would
produce a bright prompt emission followed by a somewhat dim afterglow.
\end{abstract}
\section{Introduction}
\label{intro}

The disks of Active Galactic Nuclei (AGNs), in addition to providing fuel to their central supermassive black holes (SMBHs), are also home to stars and to the compact progenitors they leave behind. 
Stars are believed to exist in AGN disks due to two mechanisms: in-situ formation from gravitational instabilities in the outer disk (e.g. \citealt{Goodman2003,Dittmann2020}), and capture from the nuclear star cluster surrounding the AGN (e.g. \citealt{Artymowicz1993}). 
The evolution of stars in these dense environments has been studied in detail in recent works \citep{Cantiello2021,Dittmann2021,Jermin2021}. It has been shown that, in addition to growing to very large masses, AGN stars in the pre-supernova phase are fast rotators, which makes them ideal candidates as progenitors of long gamma-ray bursts (GRBs) \citep{Jermin2021}. 

Additionally, AGN disks are also conducive to the formation of short GRBs from binary neutron star (NS) mergers and possibly from NS-BH mergers with small enough mass ratios. This is due to the easiness of compact object binary formation in AGN disks. NSs and BHs cluster in migration traps (e.g. \citealt{Bellovary2016,McKernan2020}) where binary formation via dynamical interactions is facilitated. Kinetic energy loss due to the interaction with the dense AGN disk medium further contributes to binary formation \citep{Tagawa2020}. 

In addition to long and short GRBs, the disks of AGNs are expected to
host a variety of other transients, from core-collapse supernovae
(discussed by \citealt{Grishin2021}) to Accretion Induced Collapse of
NSs \citep{Perna2021b} and White Dwarfs \citep{Zhu2021WD}, to
micro-tidal disruption events by stellar mass BHs \citep{Yang2021}. Some
of these may also be accompanied by a relativistic jet, and hence
possibly emitting $\gamma$-ray photons. Even BH-BH mergers in AGN disks
may have short GRB-like features (\citealt{Bartos2017}, see also
\citealt{Kaaz2021}),  as in the case of GW~190521, which was modeled
as due to the propagation of the binary through an AGN accretion disk
\citep{Graham2020}. Proper identification of these transients, and
hence help calibrating the number of stars and compact objects in AGN
disks, bear numerous astrophysical implications. AGN disks are a
promising channel to explain some unexpected findings of the LIGO/Virgo
data, such as BHs in the low mass gap
\citep{Abbott2020low,Yang2020,Tagawa2020} as well as in the high mass
gap \citep{Abbott2020high}, and an asymmetry in the BH spin distribution
\citep{Callister2021,McKernan2021,Wang2021a}. Additionally, they can
help constrain the neutrino background associated with relativistic
sources \citep{Zhu2021nu,Fasano2021}.  As the prompt $\gamma$-ray
emission is generally followed by longer wavelength radiation, it would
contribute to the AGN variability in the optical and infrared
\citep{Wang2022}. With the upcoming Vera Rubin observatory some of these
transients will be observed, and hence it becomes especially important
to recognize them in association with their prompt $\gamma$-ray
emission.

The environment of AGN disks, due to its high density, can significantly change the appearance of a transient. Some of the general key features were discussed in \citet{Perna2021}. They showed how, depending on the  mass of the SMBH (and hence of the AGN disk), and the location of the transient within the disk, GRBs and their afterglows can appear fully normal, diffused only at early times, or completely diffused at all wavelengths and observation times. For transients emerging from the innermost regions, the jet may be chocked before emerging from the disk photosphere \citep{Zhu2021a}. 

From the gamma-ray burst point of view, the existence of different classes of events is an observational requirement. Besides the classical short and long burst classification based on their $T_{90}$ duration \citep{Kouveliotou1993}, there are at least two spectral classes within the long GRB population. Most spectra display a "tracking" behavior, in which the peak photon frequency tracks the light curve luminosity. In a subset of cases, instead, the peak frequency consistently decreases in time, irrespective of what the luminosity does (e.g., \citealt{Lu2012}). The origin of such a duality is still debated, as many of the key physical properties of the GRB prompt emission mechanism.
In the following, we will argue that bursts exploding in high density media are expected do display the hard to soft spectral behavior and suggest observations to test this scenario.
We will build on our previous work but focusing on a more in-depth study of the early phases of the jet, when the prompt emission is produced. 

The paper is organized as follows: Sec.~2 describes our numerical methods and initial conditions. The simulations results are presented in Sec.~3, and we summarize and conclude in Sec.~4.

\section{Methods}
\label{sec:methods}

\subsection{Qualitative considerations}
\label{sec:qual}

The prompt emission of gamma-ray bursts, both long and short, is predominantly attributed to either the photospheric \citep{Rees2005,Peer2006,Giannios2007,Lazzati2009,Ryde2011} or internal shock \citep{Daigne1998,Zelika2009} models, and it is likely that, at least in some cases, both mechanisms contribute to the observed radiation \citep{Guiriec2011,Toma2011}.
If a GRB occurs in low-density interstellar medium 
the standard scenario is of a relativistic outflow that releases its internal energy as gamma-ray radiation at a distance of $\sim10^{13}$~cm from the engine. Only much farther out does the interstellar medium become relevant, causing the development of a forward/reverse shock system that powers the long-lasting afterglow emission (e.g., \citealt{Meszaros1997}).  In a high density environment, on the other hand, the deceleration of the leading shell begins very early. In most cases, the photospheric emission is unaffected (\citealt{Perna2021}, see their Figure~2), but there are many locations in the disk where the deceleration takes place earlier than the internal shocks, especially for the high range of central black holes masses ($M_{\rm{BH}}>10^7$~$M_\odot$).

In this case, one can envisage that, since the deceleration of the leading shell is so sudden, the following shells catch up with the newly formed external shock very quickly. Instead of traditional internal shocks among shells, there would be shells colliding with the early formed external shock. Such collisions would take place in a small range of radii, since the shells are moving at much larger Lorentz factor than the external shock (ES), which is propagating in a high density environment and is therefore decelerating rapidly.

\begin{figure}
    \centering
    \includegraphics[width=\columnwidth]{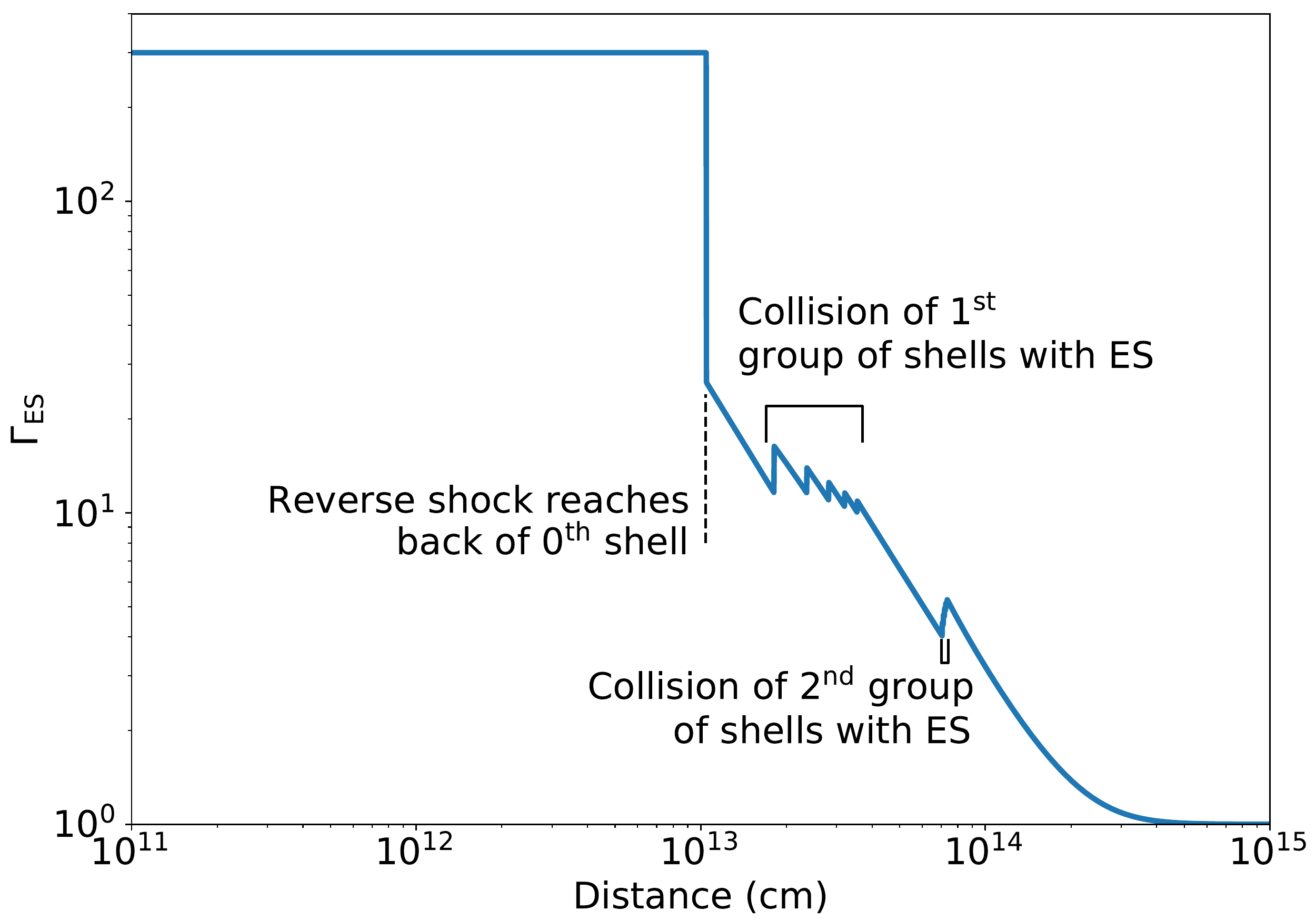}
    \caption{Dynamics of the external shocked material as a function of the distance from the central engine. The fireball is the one that gave rise to the light curve
    in Figure~\ref{fig:twopulselong}.}
    \label{fig:dynamicES}
\end{figure}

Some properties of the ensuing radiation pulses can be evaluated with simple considerations:
\begin{itemize}
    \item \emph{If the central engine produces a set of similarly spaced shells, the pulses will merge producing a single-episode prompt emission light curve}. Consider a shell that is released from the central engine at time $t_i$ after the leading shell. If it collides at a distance $R_{\rm{coll}}$ with the decelerating leading shell, the ensuing emission is seen by an observer at time $R_{\rm{coll}}/c\Gamma_0^2$, where $\Gamma_0$ is the shell's Lorentz factor before the collision. The minimum duration of the emission is instead set by the angular timescale, and is driven by the Lorentz factor of the ES: $\Delta{t}_{\rm{ang}}=R_{\rm{coll}}/c\Gamma_{\rm{ES}}$. Since the collision can happen only if $\Gamma_0>\Gamma_{\rm{ES}}$, we find that the pulse detection time (which is also the delay between the peaks from subsequent pulses), is shorter than the pulse duration. Shells emitted in a regular pattern, therefore, will produce pulses that overlap each other into a single, broad emission event. In order to observe a multi-peaked light curve, a substantial engine dead time needs to be realised.
\item \emph{The spectral evolution is hard to soft, irrespective of the light curve flux}. Given the dynamics considerations discussed above, subsequent shells collide with the leading external shock at increasing distances. The observed peak frequency of the synchrotron emission is inversely proportional to the collision distance (see, e.g., Eq. 3 in \citealt{Ghisellini2000}), creating a systematic decrease of the peak frequency with time. This is at variance with respect to the standard internal shock scenario, in which the collision radii are expected to remain at a fairly constant distance form the engine throughout the burst prompt evolution. It should be noted that the observed synchrotron peak frequency also depends on the relative Lorentz factor between the shells. An engine that produces a shell with large Lorentz factor at a late time could therefore cause a momentary increase in the peak frequency, overlaid on an overall decreasing trend. In particular, it can be expected that the peak frequency might increase after a long engine dead time, during which the external shock significantly slowed to a lower Lorentz factor.
\item  \emph{The dynamical efficiency of the prompt emission is large}. Internal shocks are known to have low dynamical efficiency \citep{Kobayashi1997,Lazzati1999}, defined as the 
ratio of the energy that is dissipated in the shock over the total energy in the outflow. This is due to the fact that the relative velocity of shells in internal shocks is mildly relativistic, since both shells are propagating at large Lorentz factor in the same direction. In the case discussed here, instead, each shell collides with a large Lorentz factor against the decelerating external shock, which is expect to be moderately relativistic ($\Gamma\sim10$, see also Figure~\ref{fig:dynamicES}). As a consequence, the dynamic efficiency is expected to be large. For a given radiative efficiency (the percentage of the dissipated energy that is radiated in electromagnetic waves), bursts in high-density media are therefore anticipated to be significantly brighter than those exploding in rarefied media.
    \item \emph{The broad-band spectrum of the prompt emission is likely to be dominated by 
    the self-synchrotron Compton process}. Self-synchrotron Compton is the process in which the synchrotron photon created by the electron population is upscattered by another electron in the same population, shifting the peak in frequency by a factor $\gamma_e^2$. The importance of this process is quantified by the Compton parameter $Y=\tau_T\gamma_e^2$, the ratio between the total energy of the Comptonized photons over the total energy of the seed synchrotron photons. For a shell hitting the external shock, we can evaluate the Compton parameter as $Y=\tau_T\gamma_e^2\simeq 1836^2 R_{\rm{ES}} n_{\rm{ISM}} \sigma_T \Gamma_{\rm{rel}}^2/\Gamma$, where $\Gamma_{\rm{rel}}$ is the Lorentz factor of the shell in the frame of the external shock and $\Gamma$ is the Lorentz factor of the merging shells in the laboratory frame. Assuming, as discussed above, that the ES does not move significantly from where it forms, we can substitute the expression for the thick-shell external shock radius \citep{Sari1995} and obtain:
    \begin{equation}
        Y \simeq 10^{-4} E_{52}^{1/4} t_{\rm{eng}}
        \Gamma_2^{-1} n_{\rm{ISM}}^{3/4} 
        \Gamma_{\rm{rel}}^2\,,
    \end{equation}
    which is typically small for low-density external media but can be substantial for typical densities of AGN accretion disks.
    
\end{itemize}

\subsection{Semi-analytical methods}

In order to validate the qualitative considerations discussed above, we carried out semi-analytical calculations of the evolution of a system of shells impacting a high-density external medium. The shells are ejected at small time intervals $\Delta{t}$  from each other\footnote{We consider an engine that ejects a shell of width $c\Delta{t}/2$, turns off for a time interval $c\Delta{t}/2$, and repeats periodically.} from a central engine embedded in a uniform medium of number density $n_{\rm{ISM}}=10^{12}$~cm${^{-3}}$, typical of the central regions of AGN accretion disks \citep{TQM,Fabj2020}. To keep the number of arbitrary assumptions at a minimum, we consider all shells ejected at regular intervals, and the thickness of each shell at ejection is set to  $\Delta{R}=c\Delta{t}/2$. We also assume that all the shells accelerate to the same asymptotic Lorentz factor $\Gamma_\infty=E/M_0 c^2$, where $E$ is the energy of each shell and $M_0$ its rest mass\footnote{Note that in traditional internal shock studies the shells are required to have different Lorentz factors to allow for internal collisions. Since in this study we concentrate on collisions of the shells with the external shock, there is no need to postulate a dispersion of Lorentz factors.}.

The first shell sweeps up the external medium. At the considered  external densities and for typical shell thicknesses and energies, the shell deceleration takes place in the thick shell regime \citep{Sari1995}. In this case the external shock radius is identified as the distance at which the reverse shock reaches the back of the shell. It reads:
\begin{eqnarray}
    R_{\rm{ES}}&=&\left(\frac{3E\Delta{}t}{4\pi n_{\rm{ISM}} m_p c }\right)^{1/4} \simeq \nonumber \\
    &\simeq& 4.7\times10^{13} \; E_{52}^{1/4} n_{\rm{ISM},10}^{-1/4},\Delta{t}^{1/4} \;\; {\rm cm}
\end{eqnarray}
which, for our fiducial parameters, is smaller than the internal shock radius 
\begin{equation}
    R_{\rm{IS}}=c\Delta{t}\Gamma_\infty^2 = 3\times10^{14} \; \Gamma_{\infty,2}^2 \Delta{t}\,.
\end{equation}
A comparison of these equations reveals that the external shocks develops first if 
\begin{equation}
    n_{\rm{ISM}}>6\times10^{6} \, E_{52} \, \Gamma_{\infty,2}^{-8} \, \Delta{t}^{-3} 
    \qquad\mathrm{cm}^{-3}
\end{equation}

The further evolution of the external shock is set by the amount of mass swept up according to \citep{Paczynski1993}:
\begin{equation}
    \Gamma=\frac{\Gamma_\infty+f}{\sqrt{1+2\Gamma_\infty f+f^2}}\,,
    \label{eq:gammaES}
\end{equation}
were $f$ is the fraction of the swept-up mass over the rest mass of the first shell.

\begin{figure}
\includegraphics[width=\columnwidth]{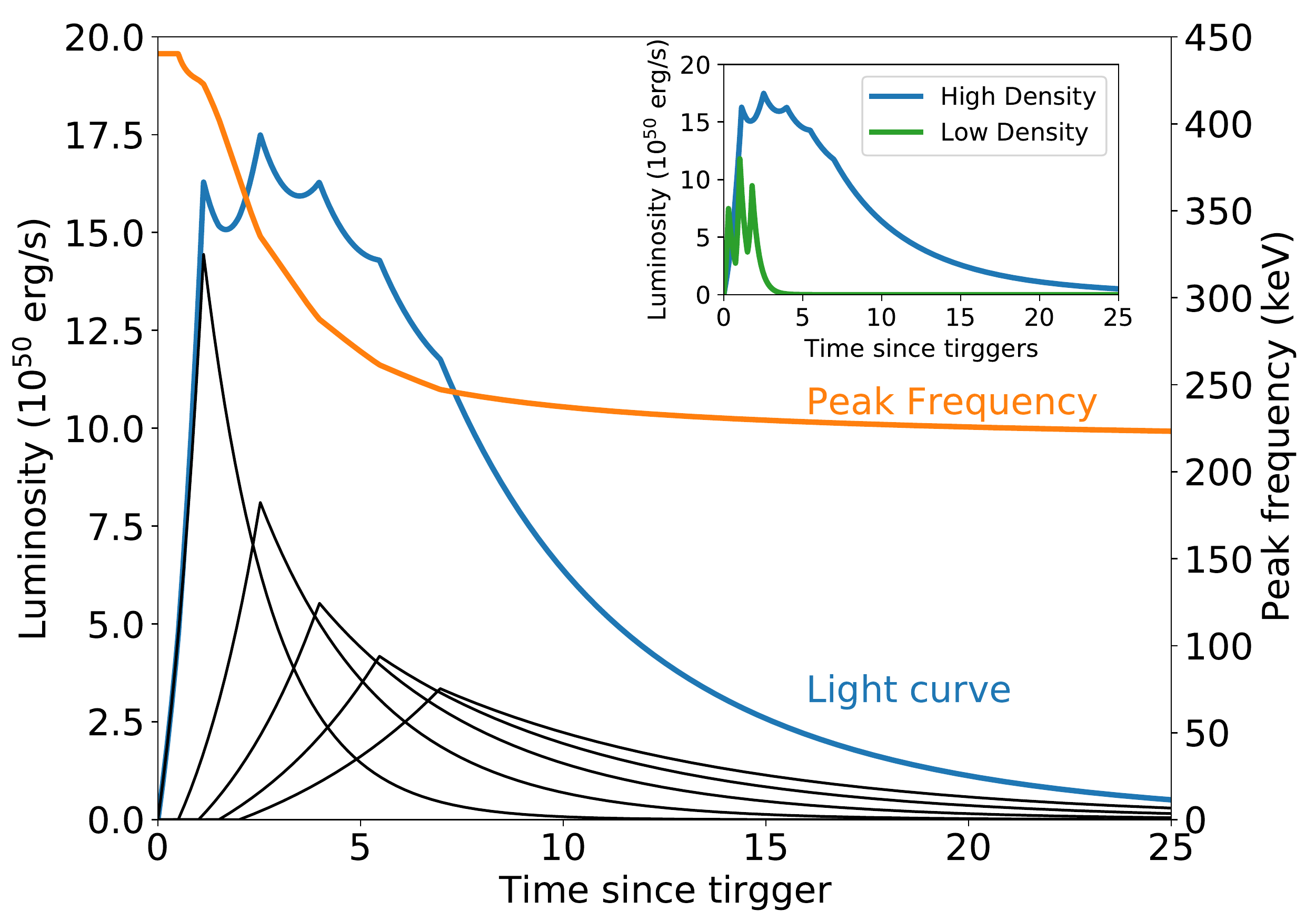}
\caption{Simulated light curve and peak frequency evolution for 
a burst exploding in a high density medium $n_{\rm{ISM}}=10^{12}$~cm$^{-3}$.
The fireball is made with an initial shell at $t=0$ followed by five equally spaced additional shells ($\Delta{t}=0.5$~s). The individual pulses from the five collisions are shown with dark thin lines, while the overall pulse profile is shown with a thick blue line. The thick orange line shows instead the evolution of the peak frequency of the spectrum (excluding SSC, see right y-axis).
 The inset shows a comparison between the prompt light curve (blue) and the prompt emission of an analogous fireball exploding in a low-density environment, and therefore powered by internal shocks.}
\label{fig:onepulse}
\end{figure}

\begin{figure}
\includegraphics[width=\columnwidth]{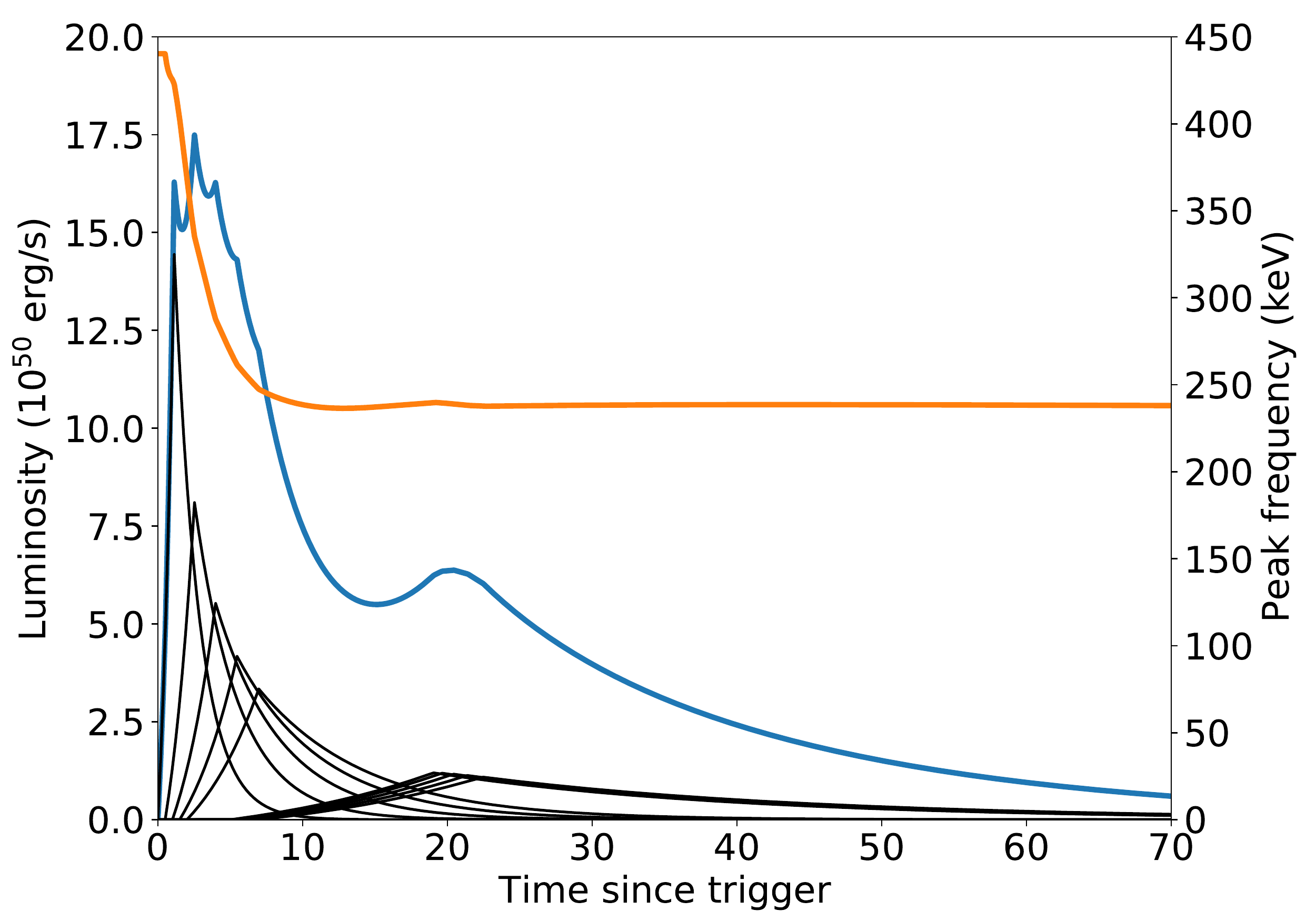}
\caption{Same as Figure~\ref{fig:onepulse}, but for a central engine that turns back on after a long pause of 5~s, ejecting another 5 shells analogous to the initial ones.}
\label{fig:twopulse}
\end{figure}

Simple kinematics equations can be used to determine the time $t_{\rm{coll}}$ and radius $R_{\rm{coll}}$ at which the second shell collides with the external shock formed by the first shell. The impact of the shell creates a prompt emission spike during a short transient in which a reverse shock propagates backward into the shell. After that, the external shock is re-energized and its velocity increases according to:
\begin{equation}
    \beta_{\rm{after-collision}} = \frac{\beta_{\rm{shell}} M_0 \Gamma_\infty+\beta_{\rm{ES}} M_{\rm{ISM}}(R_{\rm{coll}}) \Gamma_{\rm{ES}}}{ M_0 \Gamma_\infty+ M_{\rm{ISM}}(R_{\rm{coll}}) \Gamma_{\rm{ES}}}\,,
\end{equation}
where $\beta_{\rm{after-collision}}c$ is the velocity of the external shock immediately after the collision, $\beta_{\rm{shell}}c=\sqrt{1-1/\Gamma_\infty^2}c$ is the velocity of the incoming shell, $\beta_{\rm{ES}}c$ is the velocity of the external shock immediately before the collision, $M_{\rm{ISM}}(R_{\rm{coll}})$ is the mass swept-up by the external shock at the collision radius, and $\Gamma_{\rm{ES}}$ is the Lorentz factor of the external shock immediately before the collision. After the extra energy injection, the external shock settles back into its self-similar evolution, with a Lorentz factor set again by Equation~\ref{eq:gammaES}, albeit with a higher value of $M_0$. Subsequent shells have the same effect and their collision radii and kinematics can be calculated analogously. 

As discussed above, for a short time during the collision, a reverse shock forms into the shell, which can give rise to gamma-ray emission. We calculate the properties of the emission following the standard synchrotron internal shock methods (e.g., \citealt{Piran2004}), which we briefly summarize here. Electrons are accelerated to an internal Lorentz factor $\gamma_{\rm{inj}}=\Gamma_{\rm{rel}}\epsilon_e m_p/m_e$ and random, small scale magnetic field is generated with intensity $B=\sqrt{2 \epsilon_B \Gamma_{\rm{rel}} M_0 c^2/(R_{\rm{collision}}^2 \Gamma_\infty \Delta{t} \, c)}$. This gives rise to synchrotron emission with a spectrum that peaks at an observed frequency
\begin{equation}
    \nu_{\rm{syn,obs}}=\frac{2e}{3 \pi m_e \, c}\gamma_{\rm{inj}}^2 B \Gamma_{\rm{after-collision}}\,.
    \label{eq:nupk}
\end{equation}
As shown above, the conditions of the material favor a very intense self-synchrotron Compton radiation
component (see, e.g., \citealt{Marscher1985}). This has two important consequences. First, the real peak of the emission is at much higher frequencies $\nu_{\rm{SSC,obs}}=\gamma_{\rm{inj}}^2\nu_{\rm{syn,obs}}$\footnote{However, such high frequency peak might be missed in observations due to lack of instrument sensitivity and/or because the high energy photons are absorbed in photon-photon interactions either internally in the fireball of with the radiation field.}. Second, the electrons cool very rapidly, and the pulse duration is therefore expected to be driven by the fireball curvature, yielding an observed pulse with
\begin{equation}
    \Delta{t}_{\rm{pulse}} = \frac{R_{\rm{collision}}}{c\Gamma_{\rm{after-collision}}^2}\,.
\end{equation}
\vspace{0.1in}

The set of equations described here cannot be solved analytically for repeated impacts, but can be easily implemented in a computer program. In the following section we show some examples of 
predicted light curves and spectra.


\section{Results}
\label{results}

Figure~\ref{fig:dynamicES} shows the dynamical evolution of an outflow made of two shell groups impacting on a high-density external medium ($n_{\rm{ISM}}=10^{12}$~cm$^{-3}$). The graph shows the Lorentz factor of the outflow as a function of distance, with discontinuities marking the times at which shell collisions take place. In detail, the figure shows an outflow made by 11 shells. The first is emitted by the central engine at time $t=0$, followed by five that are emitted at 0.5~s intervals ($\Delta{t}=0.5$~s). Subsequently, the engine goes dormant for 20~s, and it eventually ejects a second group of shells identical to the initial ones. The dynamics of the outflow is as follows. Initially, the first shell coasts, until it develop an external forward/reverse shock system. When the reverse shock reaches the back of the first shell, a significant drop in Lorentz factor is seen in the figure, at approximately $r=10^{13}$~cm. The Lorentz factor subsequently decreases steadily, due to the accumulation of ambient material by the forward shock, until the second shell reaches the external shock and a sudden increase in Lorentz factor is observed. This sequence repeats for all subsequent shells, albeit with an overall decreasing trend. Note, however, that when the second set of shells impacts the external shock, they are so close to each other that their individual effect is difficult to discern.

Figure~\ref{fig:onepulse}  shows the light curve and peak frequency
evolution of a GRB exploding into a high density medium with
$n_{\rm{ISM}}=10^{12}$~cm$^{-3}$. The burst is made of 6 shells, all
with the same characteristics in terms of mass, energy
($E_{\rm{shell}}=10^{52}$~erg), Lorentz factor ($\Gamma_\infty=300$),
width, and separation from one another ($\Delta{t}=0.5$~s)\footnote{
The time delay among shells is reflected in the start time of the pulses
in the figures. The pulse peak time, instead, is set by the pulse
duration, which we assumed to be the angular timescale (see also the
first bullet point in Sect.~2.1).}. The figure reveals some interesting
features. As expected from the qualitative discussion in
section~\ref{sec:qual}, the individual pulses (shown with black thin
lines) are separated by a time interval that is significantly shorter
than their duration and therefore merge in a single broad envelope (the
thick blue line). In addition, later collisions give rise to longer
pulses with lower peak frequency, and therefore the overall integrated
pulse assumes a fast rise and exponential decay shape, known to occur in
GRBs\footnote{Note, however, that the so-called FRED shape can be due to
a variety of effects and is not a unique prediction of this model.}
\citep{Fenimore1996,Daigne1998}. The spectral evolution also confirms
the qualitative prediction and is overall hard to soft, irrespective of
the light curve luminosity evolution. These bursts would therefore be
outliers to the Golenetskii correlation \cite{Golenetskii1983}, which
predicts a tight correlation between luminosity and peak frequency. 
In the inset we compare the overall light curve to the prompt emission
of an analogous burst that exploded in an interstellar medium with low
density ($n_{\rm{ISM}}<10^{4}$~cm$^{-3}$; green line). In this case, the emission is powered by internal
shocks (IS) and shells were injected with random Lorentz factors between
$\gamma_{\rm{min}}=10$ and $\Gamma_{\rm{max}}=190$. The light curve (shown in green) is characterized by three individual pulses and has an overall duration comparable to the engine active time. The two fireballs contain the same energy, but the 
IS-powered light curve is much less energetic, confirming our qualitative estimate that light curves in high-density media would be more energetic. Finally, albeit not shown in the figure, the peak frequency of the second pulse of the IS-powered light curve is larger than that of the first and third pulses, making it a spectrally tracking burst.

Figure~\ref{fig:twopulse} explores a slightly more complex scenario, in which the burst's central engine goes dormant for a long time, compared to the separation between consecutive shells. In particular, the burst in the figure starts analogously to the one in Figure~\ref{fig:onepulse}, stops for 5 seconds (10 times longer than the inter-shell separation) and re-starts afterwards, by emitting another series of 5 shells identical to the initial ones. As can be seen by inspecting the figure, a second broad peak appears at about 20~s, again made by the superposition of the individual pulses. Since by that time the duration of individual pulses is much longer than the inter-pulse separation, the spectral evolution is all but gone, except for a small hint of hardening of the spectrum in correspondence with the peak of the second pulse.

\begin{figure}
    \centering
    \includegraphics[width=\columnwidth]{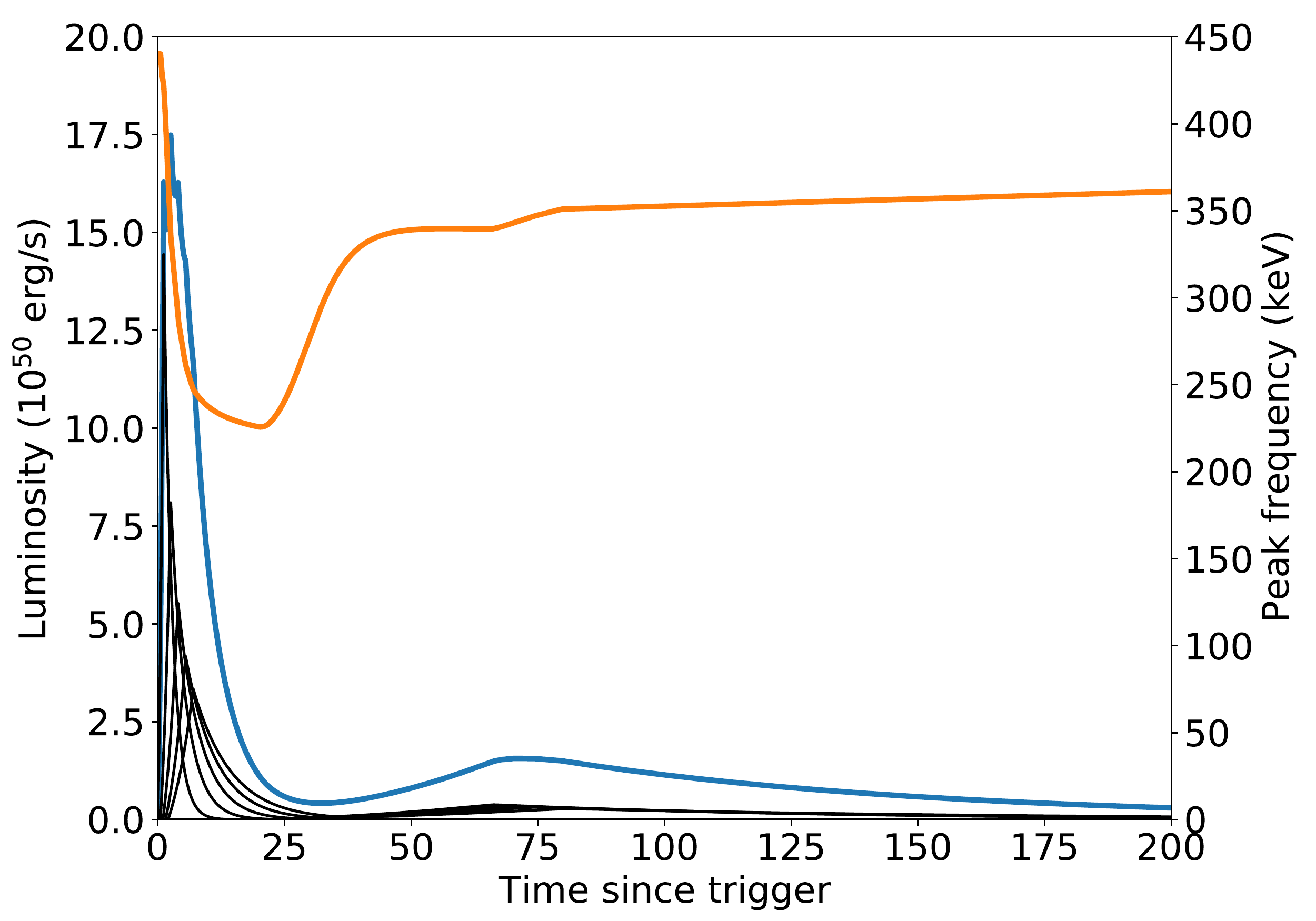}
    \caption{Same as Figures~\ref{fig:onepulse} and~\ref{fig:twopulse}, but for a central engine with an even longer pause of 20 seconds. In this case the peak frequency of the second pulse shows an increase with respect to the one at the end of the first pulse.}
    \label{fig:twopulselong}
\end{figure}

In figure~\ref{fig:twopulselong} we explore the case of an even longer dead time (20~s), so that two well-separated pulses can be seen in the light curve. This burst is the one shown in the dynamical example of Figure~\ref{fig:dynamicES}. In this case, the external shock has had enough time to slow down considerably by the time the late shells impact. Because of that, the $\gamma_{\rm{inj}}^2\propto\Gamma_{\rm{rel}}^2$ term in Eq.~\ref{eq:nupk} dominates, causing a hardening of the spectrum. For the same reason noted in the previous case, however, the spectral evolution is suppressed at late times, after the initial hardening when the second set of shells impacts the external shock.

\begin{figure}
    \centering
    \includegraphics[width=\columnwidth]{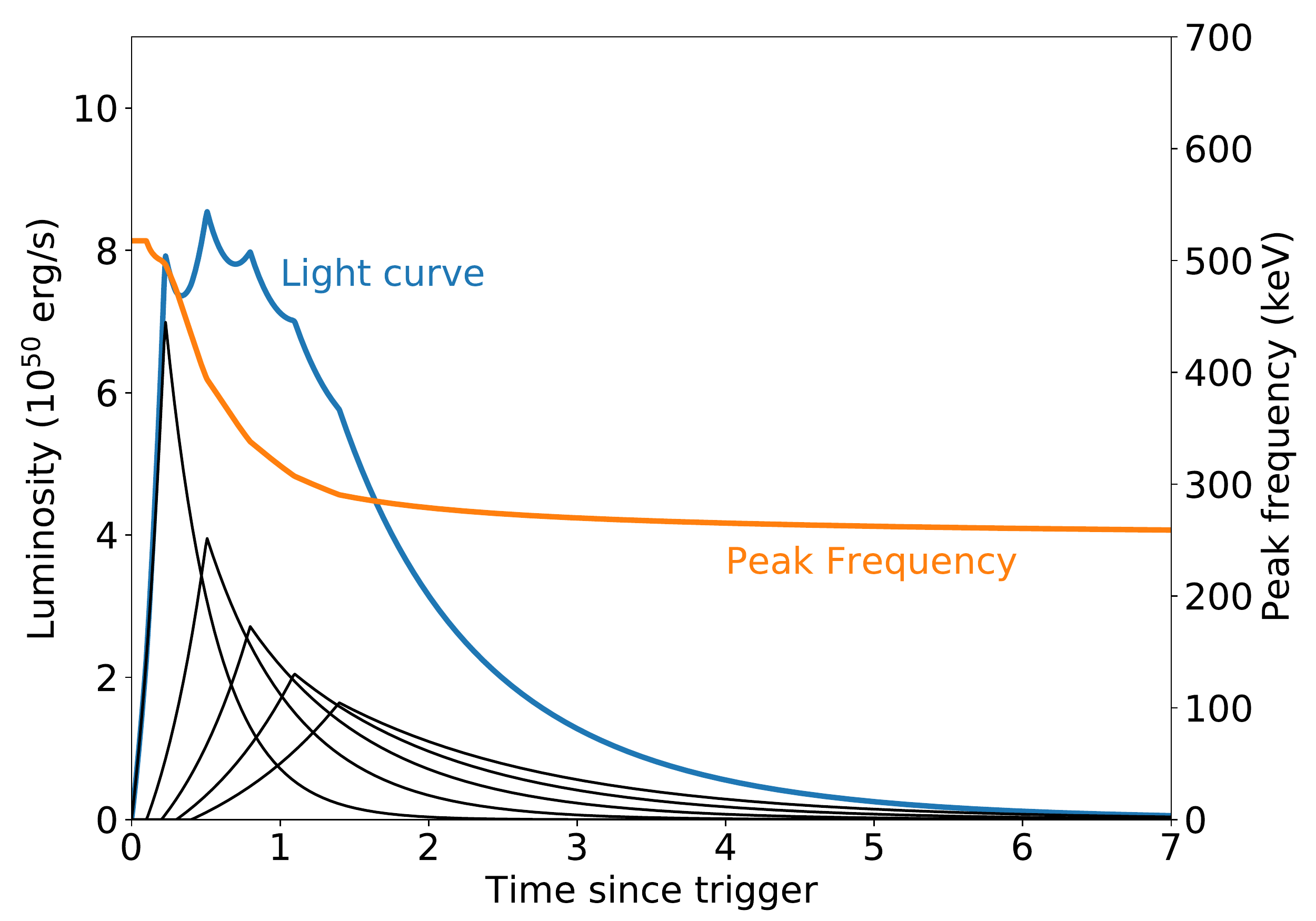}
    \caption{Same as Figures~\ref{fig:onepulse}, \ref{fig:twopulse}, and~\ref{fig:twopulselong}, but for a central engine with a short  activity that mimics a short GRB. The case shown is for an engine that releases shells at 0.1~s intervals ($\Delta{t}=0.1$~s). 
    Despite the short duration of the engine activity ($T_{\rm{eng}}=0.55$~s), the prompt light curve has a  duration of $T_{90}=3.7$~s, placing this event firmly in the long burst category.}
    \label{fig:short}
\end{figure}

An additional consideration can be made for the case of short gamma-ray bursts.
In the standard internal shock or photospheric scenarios, the duration of the prompt emission (usually characterized through the $T_{90}$ parameter) corresponds to the duration of the engine activity. In the scenario considered here, however, the duration of a pulse is set by the curvature constraint and it becomes larger for later pulses. The duration of the prompt emission is therefore potentially longer than the engine duration, and a short GRB engine may produce a burst with prompt emission lasting longer than the canonical 2~s and therefore be identified as a long burst. To check this possibility we show in Figure~\ref{fig:short} the result of our model for an engine that releases an initial shell at $t=0$ followed by five additional shells at 0.1~s intervals ($\Delta{t}=0.1$~s).
The total duration of the engine activity is therefore $T_{\rm{eng}}=0.55$~s, placing this engine in the short GRB family.
As anticipated, the resulting light curve has a much longer duration $T_{\rm{90}}=3.7$~s, which would cause the incorrect identification of this event as a long duration GRB.


\section{Discussion and conclusions}
\label{discussion}

We have presented a qualitative and a semi-analytic study of the prompt emission from gamma-ray bursts exploding in high-density media. In particular, we have focused on the consequence of the fact that, for densities larger than $\sim10^{6}$~cm$^{-3}$, 
the external shock develops before the internal shocks take place. As a consequence, the light curve is powered by a succession of shells impacting the forming external shock, each generating a strong shock system with large Lorentz factor contrast. This is different from the internal shocks case, in which mild shocks with moderate Lorentz factor contrast form due to the collisions between pairs of shells before the external shock develops.

The main conclusion of our study is the prediction that bursts from high density environments, like those that characterize the accretion disks of AGNs, should have single broad pulses with fast rise and exponential decay (FRED) shape, and that their spectral evolution should be hard to soft. Hard to soft spectral evolution is observed in GRB light curves (e.g., \citealt{Lu2012}) and its origin is still debated. In our model it is due to the fact that the FRED pulse is a superposition of shorter pulses, each of them with a decreasing peak frequency due to the increasing distance at which they are produced. While more complex light curves are possible, they require the central engine to become dormant for a period of time that is orders of magnitudes longer than the initial inter-shell separation, an occurrence that appears rare if not at all unlikely.  In addition, the dynamical efficiency of bursts exploding in high density media is larger than for traditional internal shocks, owing to the larger Lorentz factor contrast. In the cases shown in the figures, all bursts have dynamical efficiencies larger than $90\%$, compared to an efficiency of $9\%$ for the internal shock comparison in the inset of Figure~\ref{fig:onepulse}. Finally, the prompt emission duration is found to exceed 2~s even for short burst engines, therefore opening the possibility that some short GRBs from high density environment have been incorrectly classified as long bursts. Some burst with long duration have indeed defied classification due to the lack of a supernova component.  Two well-studied cases (GRB~060505 and GRB~060614, \citealt{Gehrels2006,Fynbo2006,DellaValla2006,GalYam2006}), however, cannot be due to bursts from AGN accretion disks due to the complexity of their prompt light curve and location outside of the center of their host galaxies. The case of GRB~111005A \citep{MichalowskI2018} is more interesting. The burst was classified as a long event with a single broad pulse, did not have an associated supernova, and was located within 1" of the center of its host galaxy. However, it was sub-energetic, and the data quality did not allow for a spectral study of the prompt emission. GRB111005A makes therefore a good candidate for a GRB powered by a binary neutron star merger within an AGN accretion disk, but the low energetics and possible small offset from the center of the galaxy deserve further investigation.

This model can be tested against a series of predictions and implications. First, hard to soft bursts should be predominantly FREDs, possibly showing evidence of variability overlaid on the overall pulse profile. Second, if an afterglow is observed from such bursts, it should show the characteristics of high external density, such as a fast evolution and a spectrum characterized by high-frequency self-absorption \cite{Wang2022}. Finally, should a precise localization be available, the burst should originate from the very center of the host galaxy. 

The study presented here is based on a series of simplifications that deserve further study and are summarized and commented upon in the following.

\begin{itemize}
    \item We have assumed that individual pulses from the collision of a shell with the external shock have no internal spectral evolution. Since electrons are expected to cool quickly, the pulse duration is expected to be due to the curvature of the shell, making this assumption reasonable. Peak evolution of order of a factor $\sim 2$ between the beginning and the end of each pulse is however likely present due to the change in the angle between the line of sight to the observer and the local outflow velocity.
    \item We have assumed that the prompt emission is entirely due to the synchrotron mechanism. However, there is convincing evidence that photospheric emission is at least contributing to the prompt emission in many bursts \citep{Rees2005,Peer2006,Giannios2007,Lazzati2009,Ryde2011,Guiriec2011,Toma2011}. Neglecting photospheric emission as a first order approximation is here justified by the fact that the shocks we consider have large Lorentz factor contrast and therefore are very efficient, compared to internal shocks that are expected to have small radiative efficiency \citep{Kobayashi1997,Lazzati1999}.
    \item With the aim of minimizing any arbitrary choice of parameters, we have considered an outflow made of identical shells. While this choice allows for robust conclusions and predictions, it may overlook important features of the light curves and spectra. In particular, it is unlikely that an engine that has stopped producing an outflow for a long time (like those shown in Figures~\ref{fig:twopulse} and~\ref{fig:twopulselong}) will turn on at a late time producing shells with the same properties as those in the initial phase. If later shells have lower luminosity or Lorentz factors, for example, different temporal and spectral evolution would be expected. This scenario should be studied with the aid of a physical model for the outflow generation from the central engine.
   
    \item Perhaps the most important simplification we have adopted is to limit our discussion of the self-synchrotron Compton (SSC) component. In a simple scenario, the peak of the SSC emission is at a frequency that is $\gamma_{\rm{inj}}^2$ larger than the synchrotron peak. In our case, that is likely in the GeV or TeV bands, well above the frequencies at which GRBs are detected and studied. It is possible, however, that significant feedback develops if even a fraction of these high energy photons is scattered backwards in the high density ambient material (see, for example \citealt{Beloborodov2005a,Beloborodov2005b}). The ensuing spectral modifications, however, are difficult to predict in a general case and would require a more refined model of both the outflow and the external medium.  Both of these are beyond the scope of this work.
    
    \item Finally, we would like to mention that our external medium model might be oversimplified. On the one hand, the vertical structure of an AGN accretion disk is Gaussian and not constant \citep{SG,TQM}. On the other hand, there may be significant modifications due to the per-explosion stellar and binary evolution as a result of radiation pressure, winds, feedback from accretion, etc. \citep{Kimura2021,Tagawa2021,Yuan2021}.
    
\end{itemize}

\acknowledgements We would like to thank the anonymous referee for
their constructive review and suggestions that led to a significant
improvement of this paper. DL and GS acknowledge support from NSF grant
AST-1907955. RP acknowledges support by NSF award AST-2006839.

%

\bibliographystyle{aasjournal}
\bibliography{biblio}

\end{document}